\documentclass[a4paper,11pt]{article}

\usepackage{amsmath}
\usepackage{a4wide}
\usepackage{amsfonts}
\newenvironment{pf}{\textbf{Proof:}}{\hspace{\stretch{1}}$\square$}

\newtheorem{thm}{Theorem}

\newtheorem{df}{Definition}
\newtheorem{pr}{Proposition}
\newtheorem{lem}{Lemma}
\newtheorem{dthm}{Definition-Theorem}

\newcommand{\rF}{\mathcal{F}}
\newcommand{\rH}{\mathcal{H}}

\newcommand{\rK}{\mathcal{K}}
\newcommand{\rN}{\mathcal{N}}

\newcommand{\TF}{{T\Phi}}
\font\timesept=cmr7

\newcommand{\EE}{\mathbb{E}}

\newcommand{\NN}{\mathbb{N}}

\def\tr{\mathop{\rm Tr}\nolimits}
\def\O{\Omega}

\begin{document}

\title{Return to Equilibrium\\ for some Quantum Trajectories\footnote{Work supported by ANR project ``HAM-MARK" N${}^\circ$ ANR-09-BLAN-0098-01.}}

\author{St\'ephane $\textrm{ATTAL}^1$ and Cl\'ement $\textrm{PELLEGRINI}^2$}

\date{}

\maketitle
\vskip -0.5cm
\centerline{\timesept $^1$ Universit\'e de Lyon, Universit\'e de Lyon 1}
\vskip -1mm
\centerline{\timesept Institut Camille Jordan, U.M.R. 5208}
\vskip -1mm
\centerline{\timesept 21 av Claude Bernard}
\vskip -1mm
\centerline{\timesept 69622 Villeurbanne cedex, France}
\centerline{\timesept $^2$ Institut de Math\'ematiques de Toulouse }
\vskip -1mm
\centerline{\timesept Laboratoire de Statistique et de Probabilit\'e}
\vskip -1mm
\centerline{\timesept Universit\'e Paul Sabatier (Toulouse III)}
\vskip -1mm
\centerline{\timesept 31062 Toulouse Cedex 9, France}

\begin{abstract}
We consider the situation of a two-level quantum system undergoing a
continuous indirect measurement, giving rise to so-called ``quantum
trajectories''. We first describe these quantum trajectories in a
physically realistic discrete-time setup and we then justify, by going
to the continuous-time limit, the ``stochastic Schr\"odinger equation''
attached to this model. We prove return to equilibrium properties for
these equations. \end{abstract}

\section{Introduction}

Recent experiments of continuous measurement in quantum mechanics (Haroche's
team in particular), or more precisely in quantum optics, have put into evidence
the random evolution of the state of a quantum open system \cite{Har,haroche}.
In particular, one has experimentally observed "quantum jumps". These
experiments allow to study the evolution of a quantum system interacting with
some environment. They are based of the principle of indirect measurement on the
environment, in order not to perturb the evolution of the small system
\cite{barbel,Ba3,Di1,Di2,W1,W2,francesco2}.

 The stochastic models attached to these phenomenons are described by stochastic
differential equations, called ``Stochastic Schr\"odinger Equations" or also
``Belavkin Equations"
\cite{book,barbel,Ba1,Ba2,Ba3,BaZ,BaH,be2,be3,Di1,Di2,Di3,W1,W2,francesco2}.
Their solutions are called ``quantum trajectories", they describe the evolution
of the state of the small open quantum system. The stochastic differential
equations which are usually obtained in this context are of two different types.

\smallskip
Either they are of ``jump-type":
 \begin{equation}\label{intro1}
 d\rho_t=L(\rho_t)dt+\left(\frac{\mathcal{J}(\rho_t)}{Tr[\mathcal{J}(\rho_t)]}-\rho_t\right)\left(d\tilde{N}_t-\tr[\mathcal{J}(\rho_t)]dt\right)\,.
 \end{equation}
where $(\tilde{N}_t)$ is a stochastic counting process with stochatic intensity  $\int_0^t\tr[\mathcal{J}(\rho_s)]ds$.
 The operator $L$ corresponds to a Lindblad type operator and the operator  $\mathcal{J}$ describes the evolution of the system during the quantum jumps. This equation describes experiments which are called ``direct photon detection" (observation of the photon emission by an atom excited by a laser). 

\smallskip
Or it can be an equation of diffusive type:
\begin{equation}\label{intro2}
d\rho_t=L(\rho_t)\,dt+\Big(C\rho_t+\rho_tC^\star-\textrm{Tr}[(C+C^\star)\rho_t]\rho_t\Big)\,dW_t,
\end{equation} 
where $(W_t)$ is a standard Brownian motion. In quantum optics, this equation describes experiments called ``Heterodyne or Homodyne detection''.

\smallskip
In the usual literature, obtaining and justifying rigorously these equations makes use of Quantum Filtering Theory \cite{Ba3,Ba1,be3,BGM}. It is the quantum probability version
 of the usual filtering technics, it makes use of fine Quantum Stochastic Calculus and heavy von Neumann Algebra Theory. Others approaches are based on classical probability and use of Instrumental Process and notion of ``a Posteriori State", cf \cite{book,Ba2, infinite,BaH}.

A maybe more intuitive and more physical approach for these equations is to start from a discrete-time procedure,
 that is, repeated quantum interactions with measurement of the environment (\cite{AP1}).
 One can then obtain the stochastic Schr\"odinger equations by passing to the limit to a continuous-time model \cite{Pe1,Pe2}.  

In this article, we come back and apply results obtained in  \cite{Pe1} and \cite{Pe2}, in which Belavkin equations are obtained with this approach. Here, we obtain the description of the stochastic Schr\"odinger equations for a two level system in contact with a spin chain. We adapt the result of \cite{Pe1} and \cite{Pe2} in order to describe the quantum trajectories in terms of wave function (in \cite{Pe1} and \cite{Pe2}, the stochastic equations for the evolution of density matrices have been derived from the approximation procedure). Next, for a special model, we show a property of return to equilibrium of the solution.

\section{Discrete-Time Quantum Trajectories}

In this section we describe the physical model and the mathematical
setup of indirect repeated quantum measurements.  We describe the evolution of the small
system undergoing successive measurements through the ``discrete
quantum trajectories''.

\subsection{Repeated Quantum Interactions}

The physical situation is the following. A
quantum system, with state space $\rH_S$ (often called \emph{small
system} for it is in general finite-dimensional and small compared to
the environment) is undergoing repeated
interactions with a chain of quantum systems $\otimes_{\NN^*}\rH$. This is to say that we consider an environment which is made up of a sequence of
identical copies of a quantum system, each with state space
$\rH$. Each piece $\rH$ of the environment is going to interact, one
after the other, with the small system $\rH_S$. This interaction lasts
for a time duration $\tau$ and is driven by a total Hamiltonian
$H_{\rm tot}$ on $\rH_S\otimes\rH$. Hence, each interaction is
described by the unitary operator
$$
U=e^{-i\tau H_{\rm tot}}
$$
on $\rH_S\otimes\rH$.
 In the Schr\"odinger
picture, if $\rho$ denotes any initial state on the tensor product
$\rH_S\otimes\rH$ then the
evolution of the state after this interaction is given by:
$$
\rho\mapsto U\,\rho\, U^*\,.
$$
After this  interaction, the systems $\rH_S$ and $\rH$ stop
interacting together, the system $\rH_S$ comes to meet a second
copy of  $\mathcal{H}$ and they  interact together in the same way as
before (that is, with following the same unitary operator $U$). And so on ... the small system interacts repeatedly with each of the independent copies of $\mathcal{H}.$

Let us develop the mathematical framework which allows to describe
these repeated quantum interactions. We follow the setup of the
article
\cite{AP1}, in which these models and their continuous limit were
first introduced.

The state space describing the whole game is
\begin{equation}
{\Gamma}=\mathcal{H}_S\otimes\bigotimes_{k\in\NN^*}\mathcal{H}_k\,,
\end{equation}
where each $\rH_k$ is a copy of the Hilbert space $\rH$. We have to be clear about what the above
\emph{countable} tensor product means:
$$
\TF=\bigotimes_{k\in\NN^*}\mathcal{H}_k\,.
$$
Recall that a countable tensor product of Hilbert spaces can only be defined with respect to a choice of a
particular unit vector $u_k$ in each copy $\rH_k$ (the so-called the \emph{stabilizing sequence} of the countable tensor
product). In our case, we assume that $\rH_k$ is finite dimensional and we choose an orthonormal basis
$$
\{X^i;\, i\in\rN\cup\{0\}\}
$$
where $\rN$ is a set of the form  $\{1, \ldots N\}$,
which is the same for each $\rH_k$. A particular role
is played by the vector $X^0$ which has to be considered as a
reference vector for the system $\rH$, as we choose the stabilizing sequence to be $u_k=X^0$ for all $k$. 

Denote by $X^i_k$ the basis
vector $X^i$ but leaving in the $k$-th copy $\rH_k$ of $\rH$. Then an
Hilbertian orthonormal basis of
$\TF$ is given by all the tensor products $\otimes_k v_k$ where all
the vectors $v_k$ are equal to $X^0_k$, except for a finite number of them which
might be equal to some $X^{i_k}_k$, $i_k\in\rN$. This stands for a
definition of the countable tensor product $
\TF=\otimes_{k\in\NN^*}\mathcal{H}_k$.

\smallskip
The repeated quantum interaction setup is based on two elements: the
time length $\tau$ and the Hamiltonian $H_{\rm tot}$ which  describes
each basic interaction. Consider the unitary operator $U=\exp(-i\tau
H_{\rm tot}) $ acting on $\rH_S\otimes\rH$ and consider the unitary
operator
$U_k$ on $\Gamma$  which acts  as $U$ on
$\mathcal{H}_S\otimes\mathcal{H}_k$ and which acts like the
identity operator on the other copies $\rH_{k'}$. This operator $U_k$
describes the effect of the $k$-th interaction.

The unitary operator
$$
V_k=U_k\ldots U_1
$$
describes the effect of the $k$ first interactions. Indeed, if $\rho$
is any initial state on $\Gamma$, then
 $$ V_k\,\rho\, V^*_k$$
is the state of the whole system (small system + environment) after
$k$ interactions.

\smallskip
Define the elementary operators $a^i_j$, $i,j\in\rN\cap\{0\}$ on $\rH$
by
$$
a^i_j\, X^k=\delta_{i,k}\,X^j\,.
$$
It is useful for further computations to notice that in Dirac notation $a_j^i=\vert X^j\rangle\langle X^i\vert$. We denote by $a^i_j(n)$ their natural ampliation to $\TF$ acting on
the $n$-th copy of $\rH$ only.

Clearly, $U$ can
always be written as
$$
U=\sum_{i,j\in\rN\cup\{0\}} U^i_j\otimes a^i_j
$$
for some operators $U^i_j$ on $\rH_0$ such that:
$$\sum_{k\in\rN\cup\{0\}} {(U^k_i)}^*\, U^k_j=\sum_{k\in\rN\cup\{0\}}  U^k_j\,{(U^k_i)}^*=\delta_{i,j}\,I.$$
With this representation for $U$, it is clear that the operator $U_n$,
representing the $n$-th interaction, is given by
$$
U_n=\sum_{i,j\in\rN\cup\{0\}} U^i_j\otimes a^i_j(n)\,.
$$
With these notations, the sequence ${(V_n)}$ of
unitary operators describing the $n$ first repeated interactions can be
represented as follows:
\begin{align*}
V_{n+1}&=U_{n+1}\, V_n\\
&=\sum_{i,j\in\rN\cup\{0\}} U^i_j\otimes a^i_j(n+1)V_n\,.
\end{align*}
But, inductively, the operator $V_n$ acts only on the $n$ first sites
of the chain $\TF$, whereas the operators $a^i_j(n+1)$ act on the
$(n+1)$-th site only. Hence they commute. In the following, we shall
drop the $\otimes$ symbols, identifying operators like $a^i_j(n+1)$
with $I_{\rH_0}\otimes a^i_j(n+1)$, the operator $U^i_j$ with $U^i_j\otimes I_{\TF}$, etc. This gives finally
\begin{equation}\label{E:Vn}
V_{n+1}=\sum_{i,j\in\rN\cup\{0\}} U^i_j\,V_n \, a^i_j(n+1)\,.
\end{equation}

On $\TF$, one vector plays a particular
role, the vector $$\O=\otimes_k X^0_k\,.$$ For any bounded operator $K$ on
$\Gamma$, we define the operator $\EE_0[K]$ on $\rH_S$ as the unique
operator on $\rH_S$ such that, for all trace-class operator $\rho$ on
$\rH_S$ we have
$$
\tr_{\rH_S}(\rho\,\EE_0[K])=\tr_{\Gamma}\left((\rho\otimes\vert\O><\O\vert)\,
K\right)\,.
$$
That is, $\EE_0[K]$ is the partial trace of $K$ with respect to the
state $\vert\O><\O\vert$ on $\TF$.

We then have the following fundamental action of the repeated interactions, when restricted to the small system.

\begin{thm}[cf \cite{AP1}]\label{T:CP}
The effect of the repeated
interaction dynamics when restricted to $\rH_S$ is given as
follows. For all observable $X$ on $\rH_S$,  for all $n\in\NN$, we have
$$
\EE_0[V^*_n(X\otimes I)V_n]=L^n(X)\,,
$$
where $L$ is a completely positive map on $\rH_S$ whose Krauss
decomposition is
$$
L(X)=\sum_{i\in\rN} {(U^0_i)}^*\,X\, U^0_i\,.
$$
Any (discrete) semigroup $(L^n)$ of completely positive maps can be obtained
this way.
\end{thm}

Note that the completely positive map $L$ defined above acts on
observables. It also induces a completely positive ``dual map'' $L^*$
acting on states as follows:
\begin{equation}\label{E:L*}
L^*(\rho)=\sum_{i\in\rN}  U^0_i\,\rho\,{(U^0_i)}^*
\end{equation}
and which satisfies
$$
\tr(\rho\, L(X))=\tr(L^*(\rho)\, X)
$$
for all state $\rho$ and all bounded operator $ X$ on $\rH_S$.
Recall the usual notion of partial trace
defined as follows.

\begin{dthm}
Given any state $\alpha$ on a tensor product
$\mathcal{H}\otimes\mathcal{K}$, then there exists a unique state
$\eta$ on $\mathcal{H}$ which is characterized by the property:
 $$\tr[\,\eta
\,X\,]=\tr[\alpha\, (X\otimes
I)\,]\,,$$
for all $X \in \mathcal{B}(\mathcal{H})$. The state $\eta$ is denoted
by ${\tr}_\rK(\alpha)$ and is called the \emph{partial trace of $\eta$ with
respect to $\rK$}.
\end{dthm}

With these notations we have the following result.

\begin{thm}\label{T:L*}For every state $\rho$ on $\rH_S$ and all $n\in\rN$ we have
$$
\tr_{\TF}(V_n(\rho\otimes\vert\O><\O\vert)V^*_n)={(L^*)}^n(\rho)\,.
$$
\end{thm}
\begin{pf} We have, for all $X$ bounded operator on $\rH_S$,
\begin{align*}
\tr({(L^*)}^n(\rho)\, X)&=\tr(\rho \,L^n(X))\\
&=\tr\left(\rho\, \EE_0[V^*_n(X\otimes I)V_n]\right)\\
&=\tr\left((\rho\otimes\vert\O><\O\vert)\,V^*_n(X\otimes
I)V_n\right)\\
&=\tr\left(V_n(\rho\otimes\vert\O><\O\vert)V^*_n\,(X\otimes
I)\right)\\
&=\tr\left(\tr_{\TF}\left(V_n(\rho\otimes\vert\O><\O\vert)V^*_n\right)\,X\right)\,.
\end{align*}
This proves the announced result.
\end{pf}

\subsection{Repeated Quantum Measurements}

We now somehow consider a more complicated procedure. After each
interaction is finished, the piece $\rH_k$ of environment which has
just finished to interact with $\rH_S$ is undergoing a quantum
measurement of one of its observables. The random result of this
quantum measurement will give some information on the state of the
whole system and in particular on the state of $\rH_S$. The so-called
quantum trajectory is the random process we obtain this way, by
looking at the knowledge we have of the state of $\rH_S$ after each
measurement.

Let $A$ be any
observable on $\mathcal{H}$, with spectral decomposition
$$
A=\sum_{j=1}^{p}\lambda_j P_j\,,
$$
the $\lambda_j$'s being the eigenvalues, the $P_j$'s being the
eigenprojectors. We consider the natural
ampliations of $A$ which defines an observable on $\mathbf{\Gamma}$ by
making $A$ acting on the $k$-th site $\rH_k$ only:
\begin{align*}
A^k&=\bigotimes_{j=0}^{k-1}I\otimes A\otimes\bigotimes_{j\geq k+1}I\\
&=\bigotimes_{j=0}^{k-1}I\otimes \left(\sum_{j=1}^{p}\lambda_j
P_j\right)\otimes\bigotimes_{j\geq k+1}I\\
&=\sum_{j=1}^p \lambda_jP_j^k\,,
\end{align*}
with obvious notations.

As a consequence, if $\rho$ is the state of $\mathbf{\Gamma}$ then a
quantum measurement of the observable $A^k$ gives the values $\lambda_j$ with
probability:
 $$P[\textrm{to observe}\,\,\lambda_j]=Tr[\,\rho\,P^k_j\,],\,\,\,\,j\in\{1,\ldots,p\}\,.$$
 If we have observed the eigenvalue $\lambda_j$ for the observable $A^k$, the new state of the system is
 $$
\rho_j=\frac{P^k_j\,\rho\,P^k_j }{Tr[\,\rho\,P^k_j\,]}\,.
$$ This principle is the so-called ``von Neumann projection postulate''. Now, if we
 perform another measurement of the observable $A^k$ we
obtain $P[\textrm{to observe}\,\,\lambda_j]=1$. As a consequence, a
naive repeated measurement  operation gives no information on the
evolution of the system. The repeated measurement procedure has to be
combined with the
repeated interaction procedure in order to give non-trivial
informations on the behavior of the system.

 The
quantum repeated measurement principle is the combination of the
measurement principle and the repeated quantum interactions.
Physically, this means that each copy $\rH_k$ of $\rH$ interacts with $\rH_S$ and
we perform a measurement of $A^k$ on $\rH_k$ after it has  interacted
with $\rH_S$. After each
measurement we have a new (random) state of the whole system, given by
the projection postulate.
This is the so-called \emph{discrete quantum trajectory}.

More precisely, the initial state on
 $\mathbf{\Gamma}$ is chosen to be of the form
 $$\mu=\rho\otimes\bigotimes_{j\geq 1}\eta_j,$$
 where $\rho$ is any state on $\mathcal{H}_0$ and each $\eta_i=\eta$
 is a
reference state on $\mathcal{H}$. We denote by $\mu_k$ the state
representing the new state after the $k$ first interactions, that
is,
$$
\mu_k=V_k\,\mu\, V_k^*\,.
$$
 Let us now define the probabilistic framework in order to describe the effect of the
successive measurements. We put $\Omega=\{1,\ldots,p\}$ and on
$\Omega^\mathbb{N}$ we define the cylinders of size $k$:
 $$\Lambda_{i_1,\ldots,i_k}=\{\omega\in\Omega^\mathbb{N}/\omega_1=i_1,\ldots,\omega_k=i_k
\}\,.$$
 We endow $\Omega^\mathbb{N}$ with the $\sigma$-algebra $\rF$ generated by all these sets,
this is the \emph{cylinder $\sigma$-algebra}.
 Note that for all $j$, the unitary operator $U_j$ commutes with all
 the projectors $P^k_i$ such that $k\not =j$. Hence, the state of the
 system after $k$ interactions and $k$ measurements which have given
 the respective values $\lambda_{i_1}, \ldots, \lambda_{i_k}$ is (up
 to normalization by the trace)
\begin{align*}
P^k_{i_k}\,U_k\,\ldots\, P^1_{i_1}U_1\,\mu\,
{(U_1)}^*\,P^1_{i_1}\,\ldots&\, {(U_k)}^*\,P^k_{i_k}=\\
&=P^k_{i_k}\,\ldots\,
P^1_{i_1}\,U_k\,\ldots\,
U_1\,\mu\,{(U_1)}^*\,\ldots\,{(U_k)}^*\,P^1_{i_1}\,\ldots\, P^k_{i_k}\\
&=P^k_{i_k}\,\ldots\,
P^1_{i_1}\,\mu_k\,P^1_{i_1}\,\ldots\, P^k_{i_k}\,.
\end{align*}

We denote by $
\tilde{\mu}(i_1,\ldots,i_k)$ the quantity $$P^k_{i_k}\,\ldots\,
P^1_{i_1}\,\mu_k\,P^1_{i_1}\,\ldots\, P^k_{i_k}\,.$$
By the Kolmogorov Consistency Theorem we can  define a probability
measure $P$ on $(\Omega^\mathbf{N}, \rF)$ only by specifying
 $$P[\Lambda_{i_1,\ldots,i_k}]=\tr[\tilde{\mu}(i_1,\ldots,i_k)]\,.$$

We also define a  random sequence of states on $\Gamma$ by
 $$\begin{array}{cccc}\tilde{\rho}^k(.)\ : & \Omega^{\mathbf{N}} &\longrightarrow&
\mathcal{B}(\mathbf{\Gamma})\\
 & \omega & \longmapsto & \tilde{\rho}_k(\omega_1\ldots
\omega_k)=\displaystyle{\frac{\tilde{\mu}(\omega_1\ldots
\omega_k)}{\tr[\tilde{\mu}(\omega_1\ldots \omega_k)]}}\,.
\end{array}$$
This random sequence of states is our discrete quantum trajectory
and the operator $\tilde{\rho}^k(i_1,\ldots,i_k)$ represents the
state of the system, after having observed the results $\lambda_{i_1},\ldots,\lambda_{i_k}$ for
the $k$ first measurements. This fact is made precise in the following
proposition.
\begin{pr}
Let $(\tilde{\rho}_k)$ be the above random sequence of states we
have, for all $\omega\in\Omega^{\mathbb{N}}$
$$\tilde{\rho}_{k+1}(\omega)=\frac{P^{k+1}_{\omega_{k+1}}U_{k+1}\,\tilde{\rho}_k(\omega)\,{(U_{k+1})}^*
P^{k+1}_{\omega_{k+1}}}{\tr\left[\,\tilde{\rho}_k(\omega)\,{(U_{k+1})}^*
P^{k+1}_{\omega_{k+1}}U_{k+1}\right]}\,.$$
\end{pr}

This proposition is obvious but summarizes the quantum repeated
measurement principle. The sequence $\tilde{\rho}_k$ is the
quantum trajectory, showing up the effect of the successive
measurements on $\mathbf{\Gamma}$. The following theorem is an easy
consequence of the previous proposition.

\begin{thm}\label{T:rho_global}
 The sequence $(\tilde{\rho}^n)_n$ is a Markov chain, valued in the
set of states
  of $\Gamma$. It is described as follows:
\begin{eqnarray*}
P\left[\tilde{\rho}^{n+1}=\mu\,\vert\,\tilde{\rho}^n=\theta_n,\ldots,\tilde{\rho}^0=\theta_0\right]
=P\left[\tilde{\rho}^{n+1}=\mu\,\vert\,\tilde{\rho}^n=\theta_n\right]\,.
\end{eqnarray*}If $\tilde{\rho}^n=\theta_n$ then $\tilde{\rho}^{n+1}$ takes one
of the values:
$$ \frac{
P_i^{n+1}U_{n+1}\,\theta_n\,{(U_{n+1})}^*
P_i^{n+1}}{\tr\left[U_{n+1}\,\theta_n\,{(U_{n+1})}^*
P_i^{n+1}\,\right]}\,,\,\,\,\,i=1,\ldots,p\,,$$ with probability
$\tr\left[U_{n+1}\,\theta_n\,{(U_{n+1})}^*\,P_i^{n+1}\,\right].$
\end{thm}

The most interesting behavior of the Markov chain of states above is
obtained when one restricts it to the small system $\rH_S$.
This way we  obtain a quantum  trajectory on the states of $\rH_S$ by
considering the sequence of random states on $\rH_S$:
\begin{equation}\label{TrPart}
\rho_n(\omega)=\tr_\TF(\tilde{\rho}_n(\omega))\,.
\end{equation}
This defines a sequence of state on $\mathcal{H}_S$ which contains
the "partial" information given by the measurement and we have the
following theorem which completely describes the behavior of this
random sequence.

\begin{thm}\label{RaS}
The random sequence defined by formula $(\ref{TrPart})$ is a
Markov chain with values in the set of states on $\mathcal{H}_S$.
If $\rho_n=\chi_n$ then $\rho_{n+1}$ takes one of the values:
$$\frac{\tr_\rH\left[(I\otimes P_i)\,U(\chi_n\otimes\eta)U^*\,(I\otimes
P_i)\right]}{\tr[\,U(\chi_n\otimes\eta)U^{*}\,(I\otimes P_i)]}\,,
\,\,\,\,i=1,\ldots, p\,,$$ with probability
$\tr\left[U(\chi_n\otimes\eta)U^*\,(I\otimes P_i)\right]$.

\smallskip
The expectation of $\rho_n$ satisfies
$$
\EE[\rho_n]={(L^*)}^n(\rho_0)\,,
$$
where $L^*$ is the completely positive map described in Theorem \ref{T:L*}.
\end{thm}

\begin{pf}
Assume, by induction,  that $\rho_n$ is given. This means that $\tr_{\TF}(\tilde{\rho}_n)=\rho_n$. The next
step of the quantum measurement gives (by Theorem \ref{T:rho_global})
$$
\tilde{\rho}_{n+1}=\frac{
P_i^{n+1}U_{n+1}\,\tilde{\rho}_n\,{(U_{n+1})}^*
P_i^{n+1}}{\tr\left[U_{n+1}\,\theta_n\,{(U_{n+1})}^*
P_i^{n+1}\,\right]}\,,$$
for some $i$.
Hence, we have to compute
$$
\tr_{\TF}(P_i^{n+1}U_{n+1}\,\tilde{\rho}_n\,{(U_{n+1})}^*
P_i^{n+1})\,.
$$
Decomposing, with obvious notations, the space $\Gamma$ into
$\rH_S\otimes\TF_{[0,n]}\otimes \rH_{n+1}\otimes
\TF_{[n+2,+\infty[}$, one notes that, by induction, the state
$\tilde{\rho}_n$ is of the form
$$
\theta_n\otimes\eta\otimes\bigotimes_{k\geq n+2}\eta
$$
where $\theta_n$ is a state on $\rH_S\otimes\TF_{[0,n]}$,
satisfying
$$
\tr_{\TF_{[0,n]}}(\theta_n)=\rho_n\,.
$$
Hence, for all $X$, bounded operator on $\rH_S$, we have
\begin{align}
\tr\big(\tr_{\TF}&(P_i^{n+1}U_{n+1}\,\tilde{\rho}_n\,{(U_{n+1})}^*
P_i^{n+1})\,X\big)=\nonumber\\
&=\tr\left((P_i^{n+1}U_{n+1}\,\tilde{\rho}_n\,{(U_{n+1})}^*
P_i^{n+1})\,(X\otimes I)\right)\nonumber\\
&=\tr\left(
U_{n+1}\,\tilde{\rho}_n\,{(U_{n+1})}^*
\,(X\otimes I_{[0,n]}\otimes P_i^{n+1}\otimes I_{[n+2,+\infty[})\right)\nonumber\\
&=\tr\left(U_{n+1}\,\left(
\theta_n\otimes \eta\otimes\bigotimes_{k\geq n+2}\eta\right)\,{(U_{n+1})}^*
\,(X\otimes I_{[0,n]}\otimes P_i^{n+1}\otimes
I_{[n+2,+\infty[})\right)\nonumber\\
&=\tr\left(\left(
\,\theta_n\otimes \eta\right)\,{(U_{n+1})}^*
(X\otimes I_{[0,n]}\otimes P_i^{n+1})U_{n+1}\right)\,.\label{E:star}
\end{align}
But $U_{n+1}$ acts only on $\rH_S\otimes \rH_{n+1}$, hence the
operator ${(U_{n+1})}^*
(X\otimes I_{[0,n]}\otimes P_i^{n+1})U_{n+1}$ is an operator on
$\rH_S\otimes \rH_{n+1}\otimes\TF_{[0,n]}$ (note the interchange of
space, for simplicity of the notations) which is of the form
$$
({(U_{n+1})}^*
(X\otimes P_i^{n+1})U_{n+1})\otimes I_{[0,n]}\,.
$$
Hence, the quantity (\ref{E:star}) is equal to
$$
\tr\left(\tr_{\TF_{[0,n]}}\left(
\,\theta_n\otimes \eta\right)\,{(U_{n+1})}^*
(X\otimes  P_i^{n+1})U_{n+1}\right)\,.
$$
But $\tr_{\Gamma_[0,n]}\left(
\,\theta_n\otimes \eta\right)$ is equal to
$
\tr_{\Gamma_[0,n]}\left(
\,\theta_n\right)=\rho_n\otimes\eta\,.
$
This gives finally
\begin{align*}
\tr\big(\tr_{\TF}&(P_i^{n+1}U_{n+1}\,\tilde{\rho}_n\,{(U_{n+1})}^*
P_i^{n+1})\,X\big)=\\
&=\tr\big((P_i^{n+1}U_{n+1}\,(\rho_n\otimes\eta)\,{(U_{n+1})}^*
P_i^{n+1})\,X\big)\,.
\end{align*}
But in this expression, the index $n+1$ plays no more role and the
expression above may as well be written
$$
\tr\big((P_iU\,(\rho_n\otimes\eta)\,{(U)}^*
P_i)\,X\big)
$$
on $\rH_S\otimes\rH$. This proves the first part of the theorem.

Let us check, the one concerning the expectation of $\rho_n$.
Note that the expectation of $\rho_1$ is equal to
\begin{align*}
\EE[\rho_1]&=\sum_{i=1}^pP(\{i\})\,\frac{\tr_\rH(P_{i}\,U(\rho_0\otimes
\eta)U^*P_{i})}{P(\{i\})}\\
&=\sum_{i=1}^p\tr_\rH(U(\rho_0\otimes
\eta)U^*P_{i}P_{i})\qquad\hbox{\ for }P_i\hbox{\ acts on }\rH\hbox{
only}\\
&=\tr_\rH(U(\rho_0\otimes
\eta)U^*\sum_{i=1}^pP_{i})\\
&=\tr_\rH(U(\rho_0\otimes \eta)U^*)\\
&=L^*(\rho_0)\,.
\end{align*}
We conclude easily by induction.
\end{pf}

\smallskip
Thanks to the above description we can express a discrete-time evolution equation for the quantum trajectories. Let us put
$$
\mathcal{L}_i(\rho)=\mathbf{E}_0\left[(I\otimes P_i)\,U(\rho\otimes\eta)U^*\,(I\otimes
P_i)\right]\,,
$$
$i=1,\ldots, p$. We then have for all $\omega\in\Sigma^\mathbb{N}$ and all $k>0$:
\begin{equation}\label{E:eqdiscrete}
\rho_{k+1}(\omega)=\sum_{i=0}^p\frac{\mathcal{L}_i(\rho_k)(\omega)}{\tr[\mathcal{L}_i(\rho_k)(\omega)]}\mathbf{1}_i^{k+1}(\omega)
\end{equation}
where $\mathbf{1}_i^{k}(\omega)=\mathbf{1}_i(\omega_{k})$.

\subsection{The two-level atom model}

In this section we specialise to the case where
$\mathcal{H}_0=\mathbb{C}^2$, this is the so-called \emph{two-level atom model}.
In most of the physical applications that we have in mind, the interacting system is also of the form
$\mathcal{H}=\mathbb{C}^2$. We denote by $X_0,X_1$ an orthonormal
basis where the reference state $\eta$ is diagonal:
$$\eta=\left(\begin{array}{cc}\eta_0&0\\0&\eta_1\end{array}\right)\,.$$

Let $\Omega,X$ be any orthonormal basis of $\mathcal{H}_0$. For describing the interactions between $\rH_{0}$ and $\rH$  we choose $\Omega\otimes X_0,X\otimes
X_0,\Omega\otimes X_1,X\otimes X_1$ as an orthonormal basis of
$\mathcal{H}_0\otimes\mathcal{H}$. In such a basis, the unitary operator $U$, describing the elementary interaction, can be written as a
$2\times 2$ matrix with  coefficients being operators on
$\mathcal{H}_0$. That is, we can write $U$ as:
$$
U=\left(\begin{array}{cc}U^0_{0}&U^1_{0}\\U^0_{1}&U^1_{1}\end{array}\right)\,.
$$
Let $A$ be an observable of $\mathcal{H}$ on which we want to perform a measurement. It can
be written as $A=\lambda_0 P_0+\lambda_1 P_1$ where $\lambda_i$
are its eigenvalues and $P_i$ the corresponding eigenprojectors. Let $(P^i_{k,l})_{k,l=0,1}$ be the matrix elements of the projector $P^i$ in the basis $X_0,X_1$. Put
$$
\mathcal{L}_i(\rho)=\sum_{k,l=0,1} P^i_{k,l}\left(\eta_0 U^k_0\, \rho \,(U^l_0)^*+
\eta_1 U^k_1\,\rho\,(U^l_1)^*\right)\,.
$$
Then, if $\rho_k$ denotes the state of the  system $\rH_{S}$ after the k-th
measurement, the state $\rho_{k+1}$ takes one of the two possibles values
$$
\frac{\mathcal{L}_i(\rho_k)}{\tr[\mathcal{L}_i(\rho_k)]}.
$$
We denote
$p_{k+1}=\tr[\mathcal{L}_0(\rho_k)]$ or
$q_{k+1}=\tr[\mathcal{L}_1(\rho_k)]$ the corresponding transitions probabilities. 

\smallskip
In the rest of the paper, we concentrate on a special case of environment, where $\eta=\vert X_0\rangle\langle X_0\vert$. This situation
 corresponds to a model of heat bath at zero temperature, see \cite{AtPe} for more explanations and for positive temperature models 
(let us just stress that this choice is crucial
 and that positive temperature gives rise to completely different continuous-time behaviours).
 In this situation, the discrete quantum trajectory can be described in terms of pure states. 
More precisely, if the initial state of $\mathcal{H}_0$ is pure, the random sequence $(\rho_k)$ remains pure. This way, we can describe the evolution of $\mathcal{H}_0$ with a random sequence of vectors (wave functions).

\begin{pr}\label{P:pure}
Let $\mathcal{H}_0=\mathcal{H}=\mathbb{C}^2$ and $\eta=\vert X
_0\rangle\langle X_0\vert$. Let $(\rho_k)$ be the discrete quantum trajectories corresponding to the indirect measurement of an obervable $A$.

 If $\rho_0$ is a pure state, that is $\rho_0=\vert \psi
_0\rangle\langle \psi_0\vert$, and if the measurement is non-trivial ($A$ is not a multiple of the identity), then the state of the small system $\rho_n$ is always a pure state. In other terms, there exists a random sequence of wave functions $(\vert\psi_n\rangle)$ such that $\Vert\psi_n\Vert=1$ and such that $\rho_n=\vert\psi_n\rangle\langle\psi_n\vert$, for all $n\in\mathbb{N}$.

The sequence $(\vert\psi_n\rangle)$ is also called a \emph{discrete quantum trajectory}.
\end{pr}

\begin{pf} Since we work in $2$-dimension and since $A$ is not a multiple of identity, we have $A=\lambda_0 P_0+\lambda_1 P_1$ where the $P_i$'s are one dimensional projectors. Thus there exist two vectors $\alpha_i$, $i=0,1$ such that $P_i=\vert\alpha_i\rangle\langle\alpha_i\vert$. Now, let $\rho_0=\vert\psi_0\rangle\langle\psi_0\vert$, after the first measurement if we have observed the eignevalue $\lambda_i$, the non normalized state describing the experiment is described by
\begin{eqnarray*}
 \tilde{\rho}_1(i)&=&\mathbb{E}_0\big[I\otimes\vert\alpha_i\rangle\langle\alpha_i\vert\,\, U(\vert\psi_0\rangle\langle\psi_0\vert\otimes\vert X_0\rangle\langle X_0\vert)U^\star\,\,I\otimes\vert\alpha_i\rangle\langle\alpha_i\vert\big]\\&=&
 \sum_{k,l}\sum_{u,v}\mathbb{E}_0\left[I\otimes\vert\alpha_i\rangle\langle\alpha_i\vert\,\, \Big(U_{k}^l(\vert\psi_0\rangle\langle\psi_0\vert(U_{v}^u)^\star\otimes a_k^l\vert X_0\rangle\langle X_0\vert a_u^v\Big)\,\,I\otimes\vert\alpha_i\rangle\langle\alpha_i\vert\right]\\
 &=&\sum_{k,v}\mathbb{E}_0\left[I\otimes\vert\alpha_i\rangle\langle\alpha_i\vert\,\, \Big(U_{k}^0(\vert\psi_0\rangle\langle\psi_0\vert(U_{v}^0)^\star\otimes a_k^v\Big)\,\,I\otimes\vert\alpha_i\rangle\langle\alpha_i\vert\right]\\
 &=&\sum_{k,v}\mathbb{E}\left[\left\vert U_{k}^0\psi_0\right\rangle\left\langle U_{v}^0\psi_0\right\vert\otimes\vert\alpha_i\rangle\langle\alpha_i\vert \vert X^k\rangle\langle X^v\vert\vert\alpha_i\rangle\langle\alpha_i\vert\right]\\
 &=&\mathbb{E}\left[\left\vert \sum_k\langle\alpha_i,X^k\rangle U_{k}^0\psi_0\right\rangle\left\langle \sum_v\langle\alpha_i,X^v\rangle U_{v}^0\psi_0\right\vert\otimes\vert\alpha_i\rangle\langle\alpha_i\vert\right]\\
 &=&\left\vert \sum_k\langle\alpha_i,X^k\rangle U_{k}^0\psi_0\right\rangle\left\langle \sum_v\langle\alpha_i,X^v\rangle U_{v}^0\psi_0\right\vert\,.
\end{eqnarray*}

Now, by normalizing the vector $\sum_k\langle\alpha_i,X^k\rangle U_{k}^0\psi_0$,
 it is straightforward that we get a vector $\psi_1$ such that $\rho_1=\vert\psi_1\rangle\langle\psi_1\vert$. 
Next, by induction we can construct a sequence $\psi_n$ such that $\rho_n=\vert\psi_n\rangle\langle\psi_n\vert$ for all $n$.
\end{pf}

\bigskip
\textbf{Remark:} Such a property is at the basis of the use of ``quantum trajectory theory'' for numerical simulations of Lindblad master equations.
 Numerically, the description in terms of pure states reduces the number of parameters to control
(in comparaison with density matrices) . We recover the ``deterministic'' dynamic by taking the expectation, that is,
 $\mathbb{E}[\vert\psi_n\rangle\langle\psi_n\vert]=\mathcal{L}^n(\rho_0)$. In the continuous time version, similar properties are called
 ``unravelling\footnote{Unravelling means the description of a wave function stochastic process ($\psi_t$) such that
 $\mathbb{E}[\vert\psi_t\rangle\langle\psi_t\vert]=e^{tL}(\rho_0)$}'' 
of master equations and simulations make use of technics called ``Quantum Monte Carlo simulations''.

\smallskip
Now we wish to obtain discrete-time evolution equation which describes the stochastic evolutions
 of discrete quantum trajectories. To this end, let us introduce some notations. Let $P_0$ be the projector on $\alpha_0=(\mu,\nu)$,
 with $\Vert\alpha_0\Vert=1$ and $P_1$ the projector on $\alpha_1=(\bar\nu,-\bar\mu)$. Let define the following functions acting on vectors
 \begin{eqnarray*}\mathcal{F}_0(\vert\psi\rangle)&=&\left\vert\left[ \mu U^0_0+\nu U^0_1)\right]\psi\rangle\right.\\
 \mathcal{F}_1(\vert\psi\rangle)&=&\left\vert\left[ \bar\nu U^0_0-\bar\mu U^0_1)\right]\psi\rangle\right.\,.
 \end{eqnarray*}
 Then, the dynamic of $(\psi_n)$ can be described  by the equation
 \begin{equation}\label{E:pure}
 \vert\psi_{k+1}(\omega)\rangle=\frac{\mathcal{F}_0(\vert\psi_k(\omega)\rangle)}{\Vert \mathcal{F}_0
(\vert\psi_k(\omega)\rangle)\Vert}\mathbf{1}^{k+1}_0(\omega)+\frac{\mathcal{F}_1(\vert\psi_k(\omega)\rangle)}{\Vert \mathcal{F}_1
(\vert\psi_k(\omega)\rangle)\Vert}\mathbf{1}^{k+1}_1(\omega)\,,
 \end{equation}
 for all $\omega\in\Sigma^{\mathbb{N}}$. This equation corresponds to equation (\ref{E:eqdiscrete})
 for a two level system in terms of wave functions, i.e. the sequence $(\vert\psi_k\rangle\langle\psi_k\vert)$ satisfies
 equation (\ref{E:eqdiscrete}).
\smallskip

In the next section, we will describe the continuous time version of these equations.
 To this end, we aim at considering this discrete-time model but depending on a time-length parameter $\tau$ which we shall make tend to 0. That is, we want to pass from a discrete time interaction model to a continuous time one. This way, we shall obtain the classical Belavkin equations for quantum trajectories associated to continuous measurement. In the litterature, these equations describe a model where a two-level atom is in contact with a photon-stream.

Let $\tau=\frac{1}{n}$ be the time of interaction  between the small system and one element of the environment. Let us denote
by $U(n)$ the unitary operator associated to each interaction, it now depends of the  time of
interaction. If we had no measurement process on the environment, we will be back to the problem of going from a discrete-time repeated quantum interaction model, to a continuous time one. This problem has been completely studied in \cite{AP1}. In their article they show that, in order to get a limit evolution when $\tau$ goes to 0, we have to ask the operator $U(n)$ to satisfy certain renormalization conditions.
They have shown that the coefficients $U^i_j(n)$ must follow
well-defined time scaling in order to obtain a non-trivial limit.  Namely they
have shown that the operators $V_{[nt]}=U([nt])\ldots U_1$, $t\geq 0$, 
converges to an evolution $(V_t)_t$ which is a continuous operator process.
This process naturally satisfies a quantum Langevin equation which represents
the evolution equation of the small system + bath.

Our continuous measurement procedure does not differ much from their approach, except that we perform a measurement on the environment after each interaction.
 This is why we have to keep the same normalization for the coefficients $U^i_j(n)$ in order to get a limit. Following \cite{AP1} we assume that the total Hamiltonian, describing one elementary interaction, is of the form
$$
H_{tot}=H\otimes I+I\otimes \left(\begin{matrix}\gamma_0&0\\0&\gamma_1\end{matrix}\right)+\sqrt n\left(C\otimes a^0_1+C^*\otimes a^1_0\right)\,.
$$
That is, a typical dipole-type interaction Hamitonian with a renormalization in $\sqrt{n}$ of the field operator $a_1^0$ and $a_0^1$ in order to strengthen the force of the interaction while the time of interaction decreases.

With this Hamiltonian, it is easy to check that the coefficients of $U(n)$ are of the form
\begin{eqnarray}
U^0_0(n)&=&I+\frac 1n\left(-iH-i\gamma_0I+\frac12C^*C\right)+\circ(\frac 1n)\label{E:asympt1}\\
U^0_1(n)&=&-i\frac 1{\sqrt n} C+\circ(\frac 1n)\label{E:asympt2}\\
U^1_0(n)&=&-i\frac 1{\sqrt n} C^*+\circ(\frac 1{ n})\label{E:asympt3}\\
U^1_1(n)&=&I+\frac 1n\left(-iH-i\gamma_1I+\frac12CC^*\right)+\circ(\frac 1n)\,.\label{E:asympt4}
\end{eqnarray}

\section{Continuous Trajectories}

In this section, we implement the asymptotic expression of the coefficient $U_j^i(n)$ in the description of the quantum repeated measurements for the model of the two level atom. First we recall the convergence of discrete models to continuous models of Belavkin equations. Second, we show return to equilibrium results in this context.

\smallskip
As in shown in \cite{Pe1} and \cite{Pe2}, the continuous limit of the evolution equation is completely different, depending on wether the observable $A$ is diagonal or not in the basis
of $\eta$. The point is that the limit equation is of diffusive type when $A$ is non-diagonal and of Poisson type in the diagonal case. Inside each case, the behaviors are very comparable and differ only by some coefficients.
 This is why, it is enough here to consider only two cases:
$$
A=\left(\begin{array}{cc}0&0\\0&1\end{array}\right)=a_1^1\,,
$$
as representing the diagonal case,
or
$$
A=\left(\begin{array}{cc}0&1\\1&0\end{array}\right)=a_0^1+a_1^0\,,
$$
as representing the non-diagonal case. Here, we focus on the description of quantum trajectories in terms of pure states, while in \cite{Pe1,Pe2}, the evolution for the density matrices is considered. 

\subsection{The Poisson case}

We first start with the case $
A=a_1^1
$, for which we have
$
P_0=a_0^0.
$
 It is easy to see that we can choose $\mu=1$, $\nu=0$ for the description of the projectors $P_i$. Applying the hypothesis (\ref{E:asympt1})-(\ref{E:asympt4}), we obtain the probabilities
\begin{eqnarray*}
 p_{k+1}&=&Tr[\rho_{k}P_0]=\Vert U^0_0\vert\psi_k\rangle\Vert=1-\frac{1}{n}\,\frac{1}{2}\,\mu_k(n)+\circ\left(\frac{1}{n}\right)\,,\\
q_{k+1}&=&Tr[\rho_{k}P_1]=\Vert U^0_1\vert\psi_k\rangle\Vert=\frac{1}{n}\,\frac{1}{2}\,\mu_k(n)+\circ\left(\frac{1}{n}\right)\,,
\end{eqnarray*}
where $\mu_k(n)=\langle\psi_k,C^* C\,\psi_k\rangle$. By remarking that $\mathbf{1}^k_0=1-\mathbf{1}^k_1$, we have the following difference equation for $(\psi_k)$:
\begin{multline}
 \vert\psi_{k+1}\rangle-\vert\psi_{k}\rangle=\frac{1}{n}\left(-iH-\frac{1}{2}C^* C+\frac{1}{2}\mu_k+\circ(1)\right)\vert\psi_k\rangle+\hfill\\
 \hfill+\left(\frac{C}{\sqrt{\mu_k}}-I+\circ(1)\right)\vert\psi_k\rangle\mathbf{1}^{k+1}_1\,.\label{E:discrete_poisson}
\end{multline}

In the continuous limit, we shall see that this difference equation converges to an equation of the form
\begin{multline}
d\vert
\psi_t\rangle\,=\,\left(-iH-\frac{1}{2}\left(C^*
C+\mu_{t-}\,I\right)+\sqrt{\mu_{t-}}\,C\right)\vert \psi_{t-}\rangle \,dt+\hfill\\
\hfill+\frac{(C-\sqrt{\mu_{t-}}\,I)}{\sqrt{\mu_{t-}}}\,\vert
\psi_{t-}\rangle
\,(d\tilde{N}_t-\mu_{t-}\,dt)\label{WSJE}
\end{multline}
where $\mu_t= \langle\psi_t,C^* C\,\psi_t\rangle$ and $(\tilde{N}_t)$ is  a counting
process such that
$t\rightarrow\tilde{N}_t-\int_0^t\mu_s\,ds$ is a martingale. This is to say that $(\tilde{N}_t)$ is a counting process with stochastic intensity equal to $\int_0^t\mu_s\,ds$.
A first  problem is that equation (\ref{WSJE}) is ill-defined. Indeed, the intensity of the counting process depends on the solution itself. We need to be more precise about what we mean by a ``solution to equation (\ref{WSJE})".

\begin{df}
Let $(\Omega,{\cal F},P)$ be a probability space. A \emph{process-solution} of  the jump-equation $(\ref{WSJE})$ is a
process $(\psi_t)$ and a counting process $\tilde{N}_t$, with intensity $\int_0^t\mu_s\,ds$ where  $\mu_t= \langle\psi_t,C^* C\,\psi_t\rangle$, such that for all
t we have
\begin{multline}
\vert
\psi_t\rangle\,=\vert\psi_0\rangle+\int_0^t\left(-iH-\frac{1}{2}\left(C^*
C+\mu_{s-}I\right)+\sqrt{\mu_{s-}}C\right)\vert \psi_{s-}\rangle
ds+\hfill\\
\hfill+\int_0^t\frac{(C-\sqrt{\mu_{s-}}I)}{\sqrt{\mu_{s-}}}\vert
\psi_{s-}\rangle
(d\tilde{N}_s-\mu_{s-}ds)\,.
\end{multline}
\end{df}

This notion of solution imposes the simultaneous existence of the
process $\vert
\psi_t\rangle$ and the counting process $\tilde{N}_t$. In order to
construct such a counting process, we use a Poisson point process.

Let $(\Omega,\mathcal{F},P)$ be a probability space, on which is living a  Poisson point process $N$ on $\mathbb{R}^2$ such that the expectation of the number of points $N(\omega,B)$ lying inside a Borel set $B$ is given by
$$
\mathbf{E}[N(\,\cdot\,,B)]=\lambda(B)
$$
where $\lambda$
is the Lebesgue measure on  $\mathbb{R}^2$.

This way, $N$ defines a \emph{random measure} $B\mapsto N(\omega,B)$ on $\mathbb{R}^2$, whose volume element is denoted by $N(\omega, \,dx\, ds)$.
The following theorem shows how the
random Poisson measure is used to construct the counting process.

\begin{thm}[\cite{Pe2}]\label{process-solution}
Let $(\Omega,\mathcal{F},\mathcal{F}_{t},P)$ be a filtered
probability space on which lives a  Poisson point process $N$. The following equation
\begin{multline}
\vert\psi_t\rangle=\vert\psi_0\rangle+\int_0^t\left(-iH-\frac{1}{2}\left(C^*
C-\mu_{s-}\,I\right)\right)\vert \psi_{s-}\rangle
\,ds+\hfill\\
\hfill+\int_0^t\int_{\mathbb{R}}\frac{(C-\sqrt{\mu_{s-}}\,I)}{\sqrt{\mu_{s-}}}\vert
\psi_{s-}\rangle\,\mathbf{1}_{0\leq x\leq\mu_{s-}}\, N(dx, ds)\,.\label{E:jumpequation}
\end{multline}
admits a unique solution $(\psi_t)$ such that $\Vert\psi_t\Vert=1$ almost surely. Furthermore the process $(\vert
\psi_t\rangle)$ together with the counting process
\begin{equation}\label{pointprocess}
\tilde{N}_t=\int_0^t\int_{\mathbf{R}}\mathbf{1}_{0\leq x\leq
\mu_{s-}}\,N(dx, ds)
\end{equation}
constitute a process-solution for equation (\ref{WSJE}).
\end{thm}

Even if this theorem is just an application of the results of \cite{Pe2}, let us explain roughly how it is proved (this description will allow also to describe the return to equilibrium property in the jump case). 

In equation (\ref{E:jumpequation}) there are two parts: the ordinary differential part and the one driven by the Poisson process. Consider the collection of jumping times of the Poisson process. If there is no jump of the Poisson process $N$, we deal with an ordinary differential equation
$$
\vert
\psi_t\rangle=\vert\psi_0\rangle+\int_0^t\left(-iH-\frac{1}{2}\left(C^*
C-\mu_{s-}\,I\right)\right)\vert \psi_{s-}\rangle
\,ds\,.
$$
This equation admits a unique solution, from which we deduce the curve $t\rightarrow\mu_t$.  The first time $T_1$ when the Poisson process has a jump under this curve, the solution $\vert
\psi_t\rangle$ jumps and takes the value 
$$
\frac{C\vert \psi_{{T_1}-}\rangle}{\sqrt{\mu_{T_1-}}}\,.
$$
 After this first jump, we have a new "initial" value for $\vert\psi_t\rangle$ and the process starts again in the same way: we solve the ordinary differential equation and the solution follows it, until it meets a jump of $N$ which is bellow the curve,
 then it jumps. And so on. 
 
 \bigskip
 \textbf{Remark:} The corresponding evolution for the density matrices can be obtained by computing the stochastic differential equation for $\rho_t=\vert\psi_t\rangle\langle\psi_t\vert$. By applying the stochastic calculus rules for random Poisson measure, we get the equation
 \begin{multline}
 \rho_t=\rho_0+\int_0^t\Big(L(\rho_{s-})-C\rho_{s-}C^\star+\textrm{Tr}[C\rho_{s-}C^\star]\rho_{s-}\Big)\,ds+\hfill\\
 \hfill+\int_0^t\int_\mathbb{R}\left(\frac{C\rho_{s-}C^\star}{\textrm{Tr}[C\rho_{s-}C^\star]}-\rho_{s-}\right)\mathbf{1}_{0<x<\textrm{Tr}[C\rho_{s-}C^\star]}N(dx,ds)\,,
 \end{multline}
 where $L$ is the Lindblad operator defined by $$L(\rho)=-i[H,\rho]-\frac{1}{2}\{C^\star C,\rho\}+C\rho C^\star.$$Thus, by defining $\mathcal{J}(\rho)=C\rho C^\star$,we recover the equation (\ref{intro1}) mentionned in Introduction.

\medskip
Now that equation (\ref{E:jumpequation}) is well understood, we wish to pass  to the continuous time limit on equation (\ref{WSJE}). The appropriate topology for the convergence theorem proved in \cite{Pe2} is the Skorohod topology. Let us recall it.  For all $T>0$ we denote by $\mathcal{D}([0,T])$
the space of all c\`adl\`ag matricial process on $[0,T]$ endowed with
the Skorohod topology, that is, the topology
of the weak convergence of c\`adl\`ag processes (the convergence in
distribution).

\smallskip
The approximation result is based on the description of a quantum trajectory as the solution of a stochastic equation wich is a discretization of (\ref{E:jumpequation}). In particular, from equation (\ref{E:discrete_poisson}), we can write
\begin{eqnarray}\label{E:diff}
 \vert\psi_{[nt]}\rangle&=&\vert\psi_{0}\rangle+\sum_{k=0}^{[nt]-1}\left(\vert\psi_{k+1}\rangle-\vert\psi_k\rangle)\right)\nonumber\\
 &=&\vert\psi_{0}\rangle+\sum_{k=0}^{[nt]-1}\frac{1}{n}\left(-iH-\frac{1}{2}C^* C+\frac{1}{2}\mu_k+\circ(1)\right)\vert\psi_k\rangle+\nonumber\\
 &&+\left(\frac{C}{\sqrt{\mu_k}}-I+\circ(1)\right)\vert\psi_k\rangle\mathbf{1}^{k+1}_1\,,
\end{eqnarray}
for all $t\geq0$. An adaptation of the result of \cite{Pe2} give us the following convergence. 

\begin{thm}[\cite{Pe2}]
 Let $T$ be fixed.
Let $(\Omega,\mathcal{F},P)$ be a probability space in which lives  a Poisson point process $N$. Let $(\vert\psi_{[nt]}\rangle)_{0\leq t \leq T}$ be the discrete quantum trajectory defined by the equation $(\ref{E:diff})$. This discrete quantum trajectory converges in $\mathcal{D}([0,T])$ to the process $(\vert\tilde{\psi}_t\rangle)_{0\leq t \leq T}$ which is the unique solution  of the  stochastic differential equation
\begin{equation}
 \vert\tilde{\psi}_t\rangle=\int_0^t\left(-iH-\frac{1}{2}C^* C+\frac{1}{2}\mu_tI\right)\vert\tilde{\psi}_s\rangle ds+\int_0^t\int_{\mathbb{R}}\left(\frac{C}{\sqrt{\mu_{s-}}}-I\right)\vert\tilde{\psi}_{s-}\rangle\mathbf{1}_{0<x<\mu_{s-}}\,N(dx,ds)
\end{equation}
where $\mu_t=\langle\tilde{\psi}_t\,,\,C^* C\tilde{\psi}_t\rangle$.
\end{thm}

\subsection{The diffusive case}

We now consider the case where 
$$
A=\left(\begin{matrix}0&1
\\\\
1&0\end{matrix}\right)=\left(\begin{matrix}\frac{1}{2}&\frac{1}{2}
\\\\
\frac{1}{2}&\frac{1}{2}\end{matrix}\right)-\left(\begin{matrix}\frac{1}{2}&-\frac{1}{2}\\\\-\frac{1}{2}&\frac{1}{2}\end{matrix}\right)\,.
$$
 We have 
 $$
 P_0=\left(\begin{array}{cc}\frac{1}{2}&\frac{1}{2}\\\\\frac{1}{2}&\frac{1}{2}\end{array}\right)
 $$
  and $\mu=\nu=\frac{1}{\sqrt{2}}$. Hence, after computation we obtain:
\begin{eqnarray}
 p_{k+1}&=&Tr[\rho_k P_0]=\left\Vert\frac{1}{\sqrt{2}}(U^0_0+U^0_1)\vert\psi_k\rangle\right\Vert=\frac 12+\frac{\nu_k(n)}{\sqrt{n}}+\circ\left(\frac{1}{n}\right)\,,\\
q_{k+1}&=&Tr[\rho_k P_1]=\left\Vert\frac{1}{\sqrt{2}}(U^0_1-U^0_0)\vert\psi_k\rangle\right\Vert=\frac 12-\frac{\nu_k(n)}{\sqrt{n}}+\circ\left(\frac{1}{n}\right)\,,
\end{eqnarray}
where $\nu_k(n)={\rm Re}\langle\psi_k,C\psi_k\rangle$. 

We introduce the random variables $(X_k)$ defined by
$$
X_{k+1}=-\frac{\mathbf{1}^{k}_1-q_{k+1}}{\sqrt{p_{k+1}q_{k+1}}}\,,
$$
for all $k\geq0$. In terms of $(X_k)$, the evolution equation takes the form
\begin{multline}\label{E:discr_diff}
 \vert\psi_{k+1}\rangle-\vert\psi_k\rangle=\frac{1}{n}\left(-iH-\frac{1}{2}(C^* C-2\nu_kC+\nu_k^2I)+\circ(1)\right)\vert\psi_k\rangle+\hfill\\
 \hfill+\bigg(C-\nu_k+\circ(1)\bigg)\vert\psi_k\rangle\frac{1}{\sqrt{n}}X_{k+1}\,.
\end{multline}
The continuous diffusive equation which is the natural candidate to be the limit of equation (\ref{E:discr_diff}) is 
\begin{equation}\label{WSDE}
d\vert \psi_t\rangle\,=\left(-iH-\frac{1}{2}\left(C^*
C-2\nu_tC+\nu_t^2I\right)\right)\vert \psi_t\rangle dt+\,(C-\nu_tI)\vert \psi_t\rangle
dW_t,
\end{equation}
where $\nu_t={\rm Re}\langle\psi_t,C\psi_t\rangle$ and
$(W_t)_t$ is a one-dimensional Brownian motion.

In \cite{Pe1}, it is shown that the convergence result is highly based on the existence and uniqueness of the solution for such equation (let us stress that the coefficients are not Lipschitz). In particular, by a truncation method the following Theorem is proved in \cite{Pe1}. 

\begin{thm}[\cite{Pe1}]
Let $(\Omega,\mathcal{F},\mathcal{F}_t,P)$ be a probability space on which is defined a standard Brownian motion $(W_t)_t$.  The following stochastic
differential equation
\begin{equation}
d\vert \psi_t\rangle\,=\,\left(-iH-\frac{1}{2}\left(C^*
C-2\nu_tC+\nu_t^2I\right)\right)\vert \psi_t\rangle dt+(C-\nu_tI)\vert \psi_t\rangle\,
dW_t
\end{equation}
admits a unique solution. Furthermore, almost surely, for all $t$ we
have $\Vert\psi_t\Vert=1$.
\end{thm}

We can now consider the approximation procedure. In a similar way as the Poisson case, we can consider the difference equation
\begin{eqnarray}
 \vert\psi_{[nt]}\rangle&=&\vert\psi_{0}\rangle+\sum_{k=0}^{[nt]-1}\left(\vert\psi_{k+1}\rangle-\vert\psi_k\rangle)\right)\nonumber\\
 &=&\vert\psi_{0}\rangle+\sum_{k=0}^{[nt]-1}\frac{1}{n}\left(-iH-\frac{1}{2}(C^* C-2\nu_kC+\nu_k^2I)+\circ(1)\right)\vert\psi_k\rangle\nonumber\\
 &&+\sum_{k=0}^{[nt]-1}\bigg(C\vert\psi_k\rangle-\nu_k\vert\psi_k\rangle+\circ(1)\bigg)\frac{1}{\sqrt{n}}X_{k+1}\,.\label{nondiag}
\end{eqnarray}

We have the following result.

\begin{thm}
 Let $T$ be fixed.
Let $(\Omega,\mathcal{F},\mathcal{F}_t,P)$ be a probability space on which is defined a 
standard Brownian motion $(W_t)_t$. Let $(\vert\psi_{[nt]}\rangle)_{0\leq t \leq T}$ be the
 discrete quantum trajectory defined by the equation $(\ref{nondiag})$. This discrete quantum trajectory 
converges in $\mathcal{D}([0,T])$ for all $T$ to the process $(\vert\tilde{\psi}_t\rangle)_{0\leq t \leq T}$ 
which is the unique solution on $\Omega$ of the following stochastic differential equation:
\begin{equation}
d\vert \psi_t\rangle\,=\,(C-\nu_tI)\vert \psi_t\rangle
\,dW_t+\left(-iH-\frac{1}{2}\left(C^*
C-2\nu_tC+\nu_t^2I\right)\right)\vert \psi_t\rangle\, dt
\end{equation}where $\nu_t={\rm Re}\langle\psi_t,C\psi_t\rangle$.
\end{thm}
For sake of completeness we give some details on how to prove such a convergence.

\begin{pf}
Define the processes
\begin{eqnarray*}
\psi_{n}(t)=\vert\psi_{[nt]}\rangle,\,\,\,\,
V_n(t)=\frac{[nt]}{n},\,\,\,\,W_n(t)\,=\,\frac {1}{\sqrt{n}}\sum_{k=0}^{[nt]-1}X_{k+1}.
\end{eqnarray*}
The process $(\psi_n(t))$ satisfies
\begin{eqnarray}
\psi_n(t)&=&\int_0^t\left(-\frac{1}{2}C^*
C\psi_n(s-)+Re(\psi_n(s-),C\psi_n(s-)\rangle)C\psi_n(s-)\right)\,dV_n(s)\nonumber\\&&+\int_0^t(C\psi_n(s-)-Re(\psi_n(s-),C\psi_n(s-)\rangle)\psi_n(s-)\, dW_n(s)+\varepsilon_n(t),
\end{eqnarray}
where the terms $\varepsilon_n(t)$ corresponds of the $\circ$ terms in the equation in asymptotic form.

In order to show that  $\psi_n(t)$ converges in the Skorohod space to the solution of \begin{equation*}
 \vert\psi_t\rangle=\int_0^t\left(-\frac{1}{2}C^* C\vert\psi_s\rangle+\nu_tC\vert\psi_s\rangle\right)\,ds+\int_0^t(C\vert\psi_t
 \rangle-\nu_t\vert\psi_t\rangle)\,dW_s
\end{equation*}
we make use of the celebrated Kurtz-Protter theorem. Let us recall it. 

Recall that $[X,X]$ is defined for a semi-martingale by the
formula $[X,X]_t=X^2_t-\int_0^tX_{s-}\,dX_s$. For a finite variation process $V$ we put  $T_t(V)$ to be the total variation of $V$ on $[0,t]$. 

\begin{thm}[Kurtz-Protter, \cite{K-P}]
Suppose that $W_n$ is a martingale and $V_n$ is a finite variation
process. Assume that for each $t\geq 0$:
\begin{eqnarray*}\sup_n\mathbf{E}[[W_n,W_n]_t]<\infty\\
\sup_n\mathbf{E}[T_t(V_n)]<\infty
\end{eqnarray*} and that
$(W_n,V_n,\varepsilon_n)$ converges in distribution  to $(W,V,0)$
where W is a standard brownian motion and $V(t)=t$ for all $t$.
Let $X_n(t)$ be a process satisfying
\begin{equation*}
X_n(t)=\rho_0+\varepsilon_n(t)+\int_{0}^{t}L(X_n(s-)\,dV_n(s)
+\int_{0}^{t}\Theta(X_n(s-))\,dW_n(s)
\end{equation*}

Suppose that $X$ satisfies:
$$X_t=X_0+\int_0^tL(X_{s})\,ds+\int_0^t\Theta(X_s)\,dW_s$$
and that the solution of this stochastic differential equation is
unique. Then $X_n$ converges in distribution to $X$.
\end{thm}

In our case,  the different hypothesis above are satisfied. 
Indeed, define a filtration for the process $(W_n(.))$:
$$\mathcal{F}_{t}^{n}=\sigma(X_i,i\leq[nt]).$$ The following  is proved in [Pe1]. 

\begin{pr}
We have that $(W_n(.),\mathcal{F}^n_.)$ is a martingale. The
process $(W_n(.))$ converges to a standard Brownian motion $W_.$
when $n$ goes to infinity and
$sup_n\mathbf{E}[[W_n,W_n]_t]<\infty$.

Furthermore, we have the convergence in distribution for the
process $(W_n,V_n,\varepsilon_n)$ to $(W,V,0)$ when $n$ goes to
infinity.
\end{pr}

This proves the announced convergence.
\end{pf}

\textbf{Remark:} Using Ito rules on $\vert\psi_t\rangle\langle\psi_t\vert$, we get the equation for density matrices
$$d\rho_t=L(\rho_t)\, dt+\Big(C\rho_t+\rho_tC^\star-\textrm{Tr}[(C+C^\star)\rho_t]\rho_t\Big)\,dW_t,$$
which corresponds to the equation (\ref{intro2}) mentionned in the Introduction (the Lindblad operator $L$ has the same form as the Poisson case).

\section{Return to Equilibrium}

Now that the limit equations are established, 
we are interested into the long time behaviour of the solutions. 
We specify our investigations to the special case where 
$C=\left(\begin{array}{cc}0&1\\0&0\end{array}\right)=a_0^1$ and $H=H_R=\left(\begin{array}{cc}1&0\\0&0\end{array}\right)$.  

Writing the processes $(\psi_t)$ in terms of their coordinates; that is $(\psi_t:=(x_t,y_t))$, the Belavkin equations take the form
\begin{equation}\label{eqdifpure}
\left\{\begin{array}{ccll}x_t&=&x_0+\displaystyle{\int_0^t}\Bigg(-ix_s+{\rm Re}(\bar{x_s}y_s)y_s-\frac{1}{2}{\rm Re}(\bar{x_s}y_s)x_s\Bigg)\,ds+\int_0^t\left(y_s-{\rm Re}(\bar{x_s}y_s)x_s\right)\,dW_s\\
&&\\
y_t&=&y_0+\displaystyle{\int_0^t}\left(-\frac{1}{2}y_s-\frac{1}{2}{\rm Re}(\bar{x_s}y_s)^2y_s\right)\,ds+\int_0^t\Bigg(-{\rm Re}(\bar{x_s}y_s)y_s\Bigg)\,dW_s\end{array}\right.
\end{equation}
in the diffusive case, and
\begin{equation}\label{eqjumppure}
\left\{\begin{array}{ccll}x_t&=&x_0+\displaystyle{\int_0^t}\Bigg(-ix_s+\frac{1}{2}x_s\vert y_s\vert^2\Bigg)\,ds+\int_0^t\int_{0<x<\vert y_{s-}\vert^2}\Bigg(-x_{s-}+1\Bigg)\,N(dx,ds)\\
&&\\
y_t&=&y_0+\displaystyle{\int_0^t}\left(-\frac{1}{2}y_s+\frac{1}{2}\vert y_s\vert^2y_s\right)\,ds+\int_0^t\int_{0<x<\vert y_{s-}\vert^2}\Bigg(-y_{s-}\Bigg)\,N(dx,ds)\end{array}\right.
\end{equation} 
in the Poisson case. 

In the Poisson case, note that the intensity is  $\mu_t=\vert y_{t-}\vert^2$, so that one can restrict ourselves to the case where the jumps of the Poisson process are in between the lines  $y=1$ and $y=0$ (we have namely $\vert y_{t-}\vert^2\leq1$, for all $t$).  The function $t\rightarrow
card(N(.,[0,1]\times[0,t]))=\mathcal{N}_t$ then defines a standard
Poisson process with intensity $1$. The Poisson random measure and
the previous process generate on $[0,T]$ (for a fixed $T$) a
sequence $\{(T_i,\xi_i),i\in\{1,\ldots,\mathcal{N}_t)\}\}$
where each $T_i$ represents the jump time of $\mathcal{N}$.
Moreover the random variables $\xi_i$ are uniform random variables
on $[0,1]$. Consequently we can write our quantum trajectory as follows
\begin{equation}
\left\{\begin{array}{ccll}x_t&=&x_0+\displaystyle{\int_0^t}\Bigg(-ix_s+\frac{1}{2}x_s\vert y_s\vert^2\Bigg)\,ds+
\sum_{i=1}^{\mathcal{N}_t}\Bigg(-x_{T_i-}+1\Bigg)\mathbf{1}_{0<\zeta_i<\vert y_{T_i-}\vert^2}\\
&&\\
y_t&=&y_0+\displaystyle{\int_0^t}\left(-\frac{1}{2}y_s+\frac{1}{2}\vert y_s\vert^2y_s\right)\,ds+
\sum_{i=1}^{\mathcal{N}_t}\Bigg(-y_{T_i-}\Bigg)\mathbf{1}_{0<\zeta_i<\vert y_{T_i-}\vert^2}\end{array}\right.
\end{equation} 
\bigskip

Now, we shall investigate the large time behaviour of a solution of equation (\ref{eqdifpure}) or (\ref{eqjumppure}).
 To this end we need to notice the following lemma.

\begin{lem} Let $\left(\begin{array}{c}x_t\\y_t\end{array}\right)$ be the either solution of equation (\ref{eqdifpure}) or (\ref{eqjumppure}) starting with an initial condition of the form $\left(\begin{array}{c}x_0\\0\end{array}\right)$. Then, almost surely, we have $y_t=0$, for all $t$.
\end{lem}
\bigskip

\begin{pf}
Starting from $y_0=0$, in each case, it is easy to verify that $y_t=0$ is a particular solution for the corresponding stochastic differential equation describing the evolution of $(y_t)$. As a consequence by uniqueness of solution, almost surely, $y_t=0$ for all $t$.
\end{pf}
\bigskip

\textbf{Remark:} In both cases, if $y_t=0$ for all $t$, it is easy to see that the evolution of $(x_t)$ is given by the solution of $dx_t=-ix_t\,dt$.

 \textbf{Remark:} In terms of states, this lemma expresses that if $\psi_0=\left(\begin{array}{c}x_0\\0\end{array}\right)$, we have almost surely
$$\vert\psi_t\rangle\langle\psi_t\vert=\left(\begin{array}{cc}1&0\\0&0\end{array}\right)=\vert\Omega\rangle\langle\Omega\vert,$$
for all $t$. In other words, the state $\vert\Omega\rangle\langle\Omega\vert$ is an invariant (or stationnary state) for the stochastic dynamic of continuous measurement (let us stress that without measurement, i.e in the deterministic regime, it is easy to see that this state is already the invariant state).

Now we can make precise the result which states the return to equilibrium property. In particular we focus on the large time behaviour of the part $y_t$ and we show that this process converges to zero when $t$ goes to infinity.

\begin{pr}Let $\vert\psi_t\rangle=\left(\begin{array}{c}x_t\\y_t\end{array}\right)$ be either the solution of the
jump-equation or the solution of the diffusive equation, then we have
\begin{equation}
\vert y_t\vert^2\mathop{\longrightarrow}_{t\rightarrow\infty}^{a.s}0\,.
\end{equation}

Therefore, we have
\begin{equation}
y_t\mathop{\longrightarrow}_{t\rightarrow\infty}^{a.s}0\,
\end{equation} and the process of pure states $(\vert\psi_t\rangle\langle\psi_t\vert)$, where $\psi_t=\left(\begin{array}{c}x_t\\y_t\end{array}\right)$, for all $t$, satisfies
\begin{equation}
\vert\psi_t\rangle\langle\psi_t\vert\mathop{\longrightarrow}_{t\rightarrow\infty}^{a.s}\vert
\Omega\rangle\langle \Omega\vert\,.
\end{equation}
\end{pr}

\begin{pf}
 Let us first treat the case of the jump-equation. We need to share into two cases, if there is jumps or if there are no jumps. 
 
 In the case where there is at least one jump. At the first jumping time $T_1$ we have
$$\left(\begin{array}{c}x_{T_1}\\y_{T_1}
\end{array}\right)=\left(\begin{array}{c}x_{T_1-}\\y_{T_1-}
\end{array}\right)+\left(\begin{array}{c}-x_{T_1-}+1\\-y_{T_1-}
\end{array}\right)=\left(\begin{array}{c}1\\0
\end{array}\right)\,.
$$
 Following the description of the solution of the jump equation, the solution after $T_1$ is given by the ordinary differential part with the new initial condition $\left(\begin{array}{c}1\\0
\end{array}\right)\,$. This initial condition satisfies $y_0=0$, then by Lemma 1, we get $y_t=0$ for all $t\geq T_1$. 

If there are no jumps, this corresponds to the event $A=\{\omega\in\Omega/N(\omega,\{(s,x)\in\mathbb{R}^2/0<x<\vert y_s\vert^2\}=0\}$. In this situation, the evolution of $(y_t)$ is only given by the ordinary differential equation
$$y_t=y_0+\frac{1}{2}\int_0^t(-y_s+\vert y_s\vert^2y_s)\,ds\,.$$
We want to show that $\vert y_t\vert^2\rightarrow 0$, when $t$ goes to infinity. Derivating, we get  
$$
\frac{d}{dt}(\vert y_t\vert^2)=\frac{d}{dt}( y_t\bar{y_t})=y_t\frac{d}{dt}(\bar{y_t})+\bar{y_t}\frac{d}{dt}(y_t)=\vert y_t\vert^2(\vert y_t\vert^2-1)\,.
$$ 
By Lemma 1, if $y_0\neq 0$, we have $\vert y_t\vert^2>0$, for all $t$. In this case, dividing by $y_t^2$ we solve the equation and we get 
$$\vert y_t\vert^2=\vert y_s\vert^2\times\exp\left(-(t-s)+\int_s^t\vert y_u\vert^2du\right)\,,$$
for all $t>s$. In particular the function $t\rightarrow \vert y_t\vert^2$ is decreasing, then we get
\begin{eqnarray}\label{estimate}
\vert y_t\vert^2
&\leq&\vert y_s\vert^2\exp\left(-2(t-s)+2(t-s)\vert y_s\vert^2 \right),
\end{eqnarray}
for all $t\leq s$. Since we have $\vert x_s\vert^2+\vert y_s\vert^2=1$, for all $s$, we have $\vert y_s\vert^2\leq1$, for all $s$. 

With the estimate (\ref{estimate}), in order to conclude, we need to show that there exists $s$ such that $\vert y_s\vert^2<1$. Assume the contrary, we should have $\vert y_s\vert=1$, for all $s$. But such an event is of probability zero. Let us prove this fact. In this case the event $A$ becomes  
$$
A=\{\omega\in\Omega\,;\, N(\omega,\{(s,x)\in\mathbb{R}^2, 0<x<1,s>0\})=0\}\,.
$$ 
We claim that $A$ is of probability 0. Indeed, 
we have 
\begin{multline}
P[\{\omega\in\Omega\,;\,N(\omega,\{(s,x)\in\mathbb{R}^2,0<x<1, s>0\})=0]=\hfill\\
\hfill=\lim_nP[\{\omega\in\Omega\,;\,N(\omega,\{(s,x)\in\mathbb{R}^2/0<x<1,0<s<n\})=0\}]
\end{multline}
 and 
 $$
 P[\{\omega\in\Omega\,;\,N(\omega,\{(s,x)\in\mathbb{R}^2/0<x<1,0<s<n\})=0\}]=\exp(-n)\,.
 $$
 Hence the announced result about $A$.
 
\smallskip
As a consequence, there exist $s$ such that $\vert y_s\vert^2<1$. For this $s$, by taking the limit $t$ goes to infinity in expression (\ref{estimate}), we get $\vert y_t\vert^2\rightarrow 0$.

With the above discussion, for the jump equation, it is easy to conclude that
\begin{equation*}
y_t^2\mathop{\longrightarrow}_{t\rightarrow\infty}^{a.s}0\,.
\end{equation*}

\smallskip
Let us now treat the diffusive case. In order to prove the result we shall show first that $\vert y_t\vert^2$ converges almost surely to a random variable $u_\infty$ when $t$ goes to infinity. Second we show $u_\infty=0$ almost surely. Using Ito rules, we get
\begin{eqnarray*}
 d\vert y_t\vert^2&=&y_t\,d\bar{y_t}+\bar{y_t}\,dy_t+dy_t\,d\bar{y_t}\\
&=&-\vert y_t\vert^2\,dt-2{\rm Re}(\bar{x_t}y_t)\vert y_t\vert^2\,dW_t
\end{eqnarray*}
As a consequence we have almost surely:
\begin{equation}\label{solvert}
 y_t^2=y_s^2+\int_s^t-\vert y_u\vert^2\,du+\int_s^t-2{\rm Re}(\bar{x_u}y_u)\vert y_u\vert^2\,dW_u,
\end{equation}
for all $t>s$. Let $(\mathcal{F}_t)$ be the filtration generated by the Brownian motion, that is $\mathcal{F}_t=\sigma\{W_u,u\leq t\}$. Since $\mathbb{E}\left[\int_s^t-2{\rm Re}(\bar{x_u}y_u)\vert y_u\vert^2\,dW_u\vert\mathcal{F}_s\right]=0$, the above equation shows that
$$\mathbb{E}[\vert y_t\vert^2\vert \mathcal{F}_s]\leq\mathbb{E}[\vert y_s\vert^2].$$
This way the process $(\vert y_t\vert^2)$ is a super martingale which is bounded (for all $t$, we have $0\leq\vert y_t\vert\leq1$). Therefore, this process converges
 almost surely to a non-negative random variable $u_\infty$ when $t$ goes to infinity. In order to show that this random variable is equal to zero 
almost surely, we just have to show that $\mathbb{E}[u_\infty]=0$. To this end, from Eq. \ref{solvert} for $s=0$, we get
$$\mathbf{E}[\vert y_t\vert^2]=y_0^2+\int_{0}^t-\mathbf{E}[\vert y_s\vert^2]\,ds\,.$$
Solving the equation, we get
$$\mathbf{E}[\vert y_t\vert^2]= \vert y_0\vert^2e^{-t}\,.$$
As a consequence, we get
$$\mathbf{E}[y_t^2]\mathop{\rightarrow}_{t\rightarrow\infty}0\,.$$
Now, using the Lebesque dominated convergence Theorem, we deduce that $\mathbb{E}[u_\infty]=0$ and then $u_\infty=0$ almost surely. The proposition is then proved.
\end{pf}

\bigskip
\textbf{Remark:} In the proof, we have supposed that the initial condition is deterministic. This result can be easily genralized by assuming that the initial condition is random and the same result holds.

\bigskip
\textbf{Remark:} In Probability Theory, usually we consider invariant
 measure for stochastic process. Here the invariant measure is the Dirac 
measure on the state $\vert\Omega\rangle\langle\Omega\vert$. In \cite{BaP}, essentially for the diffusive equations,
 by assuming special conditions on the coefficients defining the stochastic differential equations,
 results for invariant measure for stochastic Schr\"odinger equations have been investigated. They established results 
where the invariant measure owns particular properties (in particular concerning the support of the measure which is absolutely
 continuous with respect to the Lebesgue measure).

\end{document}